\documentclass[12pt]{article}
\usepackage{epsfig}

\usepackage{float}
\usepackage{graphicx}

\begin{document}
\rightline{NKU-2018-SF1}
\bigskip

\newcommand{\be}{\begin{equation}}
\newcommand{\ee}{\end{equation}}
\newcommand{\noi}{\noindent}
\newcommand{\ra}{\rightarrow}
\newcommand{\bib}{\bibitem}
\newcommand{\refb}[1]{(\ref{#1})}

\newcommand{\bff}{\begin{figure}}
\newcommand{\eff}{\end{figure}}

\begin{center}
{\Large\bf   Massive gravity with Lorentz  symmetry breaking: black holes  as  heat engines}

\end{center}
\hspace{0.4cm}
\begin{center}
Sharmanthie Fernando \footnote{fernando@nku.edu}\\
{\small\it Department of Physics, Geology \& Engineering Technology}\\
{\small\it Northern Kentucky University}\\
{\small\it Highland Heights}\\
{\small\it Kentucky 41099}\\
{\small\it U.S.A.}\\

\end{center}

\begin{center}
{\bf Abstract}
\end{center}

In extended phase space, a static black hole in massive gravity is studied as a holographic heat engine. In the massive gravity theory considered, the graviton  gain a mass due to Lorentz symmetry breaking. Exact efficiency formula is obtained for a rectangle engine cycle for the black hole considered. The efficiency is computed by varying two parameters in the theory, the scalar charge Q and $\lambda$. The efficiency is compared with the Carnot efficiency for the heat engine. It is observed that when Q and $\lambda$  are increased that the efficiency for the rectangle cycle increases. When compared to the Schwarzschild AdS black hole, the efficiency  for the rectangle cycle  is larger for the Massive gravity black hole.

\hspace{0.7cm}


\section{ Introduction}

Black holes in anti-de Sitter space as a thermodynamical system has attracted lot of attention in a variety of contexts: when the negative cosmological constant is taken as the thermodynamical pressure of the black hole with the relation $ P = - \frac{\Lambda}{ 8 \pi}$, the resulting thermodynamics lead to interesting features. In this extended phase space, the first law of thermodynamics is modified by a term $V dP$ and the mass $M$ of the black hole is treated as the enthalpy rather than the internal energy $E$  of the black hole \cite{kastor} \cite{dolan}. Many black holes in the context of extended phase space has demonstrated Van der Waals type phase transitions between small and large black holes. Due to the large number of work published related to this topic we will mention few here: \cite{mann} \cite{hen1} \cite{mann3}  \cite{mann4} \cite{mann5}\cite{cao}  \cite{liu2} \cite{hendi} \cite{hendi2} \cite{mo} \cite{zhang} \cite{li} \cite{mo2} \cite{cai} \cite{cai2} \cite{azg}\cite{pou}. There is a nice review on {\it Black hole chemistry} written by Kubiznak et.al. which gives a comprehensive summary of the interesting thermodynamical features of black holes with a cosmological constant \cite{kub}.

In classical thermodynamics, there are four basic thermodynamical processes. They are isothermal, adiabatic, isobaric, and isochoric processes. In each of these processes the thermodynamical quantities temperature, entropy, pressure, and, volume are kept constant respectively. In a heat engine, a thermodynamical cycle is chosen which consist of aforementioned  processes. For example, the Carnot cycle, which has the highest efficiency, have two isothermal and two adiabatic processes in the cycle. Brayton cycle has two isobaric and two adiabatic processes. In this paper, the goal is to study a black hole in massive gravity as a heat engine. The idea that black holes could be used as a working substance in a heat engine was first presented by Johnson \cite{johnson1}. In that paper, charged black hole in AdS space in $D = 4$ was presented as an example. In an extension of that work, Johnson presented the Born-Infeld AdS black hole as a heat engine in \cite{johnson2}. In \cite{yerra}, heat engines from  dilatonic Born-Infeld black holes were analyzed. Black holes in conformal gravity as heat engine was presented by Xu et.al.  in \cite{sun}. Charged BTZ black hole in 2+1 dimensions as a heat engine was studied by Mo et.al. in \cite{mo5}. Effects of dark energy on the efficiency of heat engines of AdS black holes were analyzed in detail by Liu and Meng in \cite{meng}. Class of black holes in massive gravity as heat engines were discussed by Hendi et.al. \cite{meng2}. The first study of rotating black holes as holographic heat engines were done by Henniger et.al. \cite{mann2}. Heat engines are defined for space-times that are not black holes as well: Johnson \cite{johnson4} studied Taub-Bolt space-time as an example of a heat engine and compared with the analogous Schwarzschild black hole as a heat engine.

Massive gravity theories have become popular as a mean of explaining accelerated expansion of the universe without having to introduce a component of ``dark energy.'' There are many theories of massive gravity in the literature. There are large volume of work related to theories in massive gravity \cite{drgt1} \cite{drgt2} \cite{dgp} \cite{town}: here we will mention two nice reviews on the subject by de Rham \cite{claudia} and Hinterbichler \cite{kurt}.

In this paper we will consider a massive gravity theory where the graviton acquire a mass due to the Lorenz symmetry breaking. Here, Higgs mechanism for gravity is introduced with space-time depending scalar fields that are coupled to gravity \cite{luty}. The resulting theory will exhibit modified gravitational interactions at large scale. These models are free from ghosts and tachyonic instabilities around curved space as well as in flat space \cite{ruba} \cite{pilo1}. A  review of Lorentz violating massive gravity theory can be found in \cite{dubo} \cite{ruba2}. The details of the theory is discussed in section 2.

The paper is organized as follows: in section 2, the black hole in massive gravity is introduced. Thermodynamics and phase transitions of the massive gravity black hole are discussed in section 3. Black hole as a heat engine with a rectangle cycle is introduced in section 4 and in section 5 the efficiencies are computed. Finally the conclusion is given in section 6.


\section{ Introduction to AdS black holes in massive gravity with Lorentz symmetry breaking}

The action for the massive gravity theory considered in this paper  given by,
\be \label{action}
S = \int d^4 x \sqrt{-g } \left[ - \frac{1}{ 16 \pi}  \mathcal{R}  + \Omega^4 \mathcal{F}( X, W^{ij}) \right]
\ee
Here  $\mathcal{F}$ is a function of our scalar fields $\phi^{\mu}$. $\phi^{\mu}$ are minimally coupled to gravity by covariant derivatives. $\mathcal{F}$ depends on two combinations of the scalar fields given by  $X$ and $W$ defined in terms of the scalar fields as,
\be \label{scalar1}
X = \frac{\partial^{\mu} \phi^0 \partial_{\mu} \phi^0 }{ \Omega^4}
\ee
\be \label{scalar2}
W^{i j} = \frac{\partial^{\mu} \phi^i \partial_{\mu} \phi^j}{ \Omega^4} - \frac{\partial^{\mu} \phi^i \partial_{\mu} \phi^0  \partial^{\nu} \phi^j \partial_{\nu} \phi^0 }{ \Omega^8 X}
\ee
The constant $\Omega$ has dimensions of mass:  it is in the order of $ \sqrt{ m_g M_{pl}}$ where  $m_g$  is the graviton mass and $M_{pl}$ the Plank mass \cite{dubo} \cite{luty} \cite{ruba} \cite{pilo1}. The scalar fields $\phi^0, \phi^i$  are responsible for breaking Lorentz symmetry spontaneously when  they  acquire a vacuum expectation value. More details on the theory can be found in \cite{dubo} \cite{luty} \cite{ruba} \cite{pilo1} \cite{ruba2}. 
A class of black hole solutions to the above action was derived in \cite{tinya} and \cite{pilo}.  

\begin{equation} \label{metric}
ds^2 = - f(r) dt^2 + \frac{ dr^2}{ f(r)} + r^2 ( d \theta^2 + sin^2 \theta d \phi^2)
\end{equation}
where,
\begin{equation} \label{hr}
f(r) = 1 - \frac{ 2 M} { r} - \gamma \frac{ Q^2}{r^{\lambda}} - \frac {\Lambda r^2}{3}
\end{equation}
Here, the parameter $Q$ represents a scalar charge related to massive gravity and $\gamma = \pm 1$. The constant $\lambda$ in  is an integration constant and is positive.  $\Lambda$ is the cosmological constant. A detailed description of this black hole is given  by Fernando in \cite{fernando0}. Since for  $\lambda  <1$,  the ADM mass  become divergent, such solutions will not be considered. When $\lambda >1$, for $ r \ra \infty$  the metric approaches the Schwarzschild-AdS black hole with a finite mass $M$; hence we will choose $\lambda >1$ in the rest of the paper. \\

\noi
When $ \gamma = 1$, the space-time of the black hole is  very similar to the Schwarzschild-AdS  black hole and for  $\gamma = -1$, the geometry is similar to the   Reissner-Nordstrom-AdS  charged black hole. 

There are several works related to the black holes given above. P-V criticality and phase transitions  of masive gravity black holes with a negative cosmological constant were  presented by Fernando in \cite{fernando0} \cite{fernando1}. Thermodynamics for $\Lambda=0$ case were studied in  \cite{capela} \cite{mirza}. 
Stability and quasi normal modes were studied in  \cite{fernando2}\cite{fernando3} \cite{fernando4} \cite{and}.


\section{Thermodynamics and phase transitions}

In this section we will  derive the thermodynamical quantities and present the phase transitions that could occur in massive gravity black holes.


\subsection{ Defining the thermodynamical quantities}

\begin{figure} [H]
\begin{center}
\includegraphics{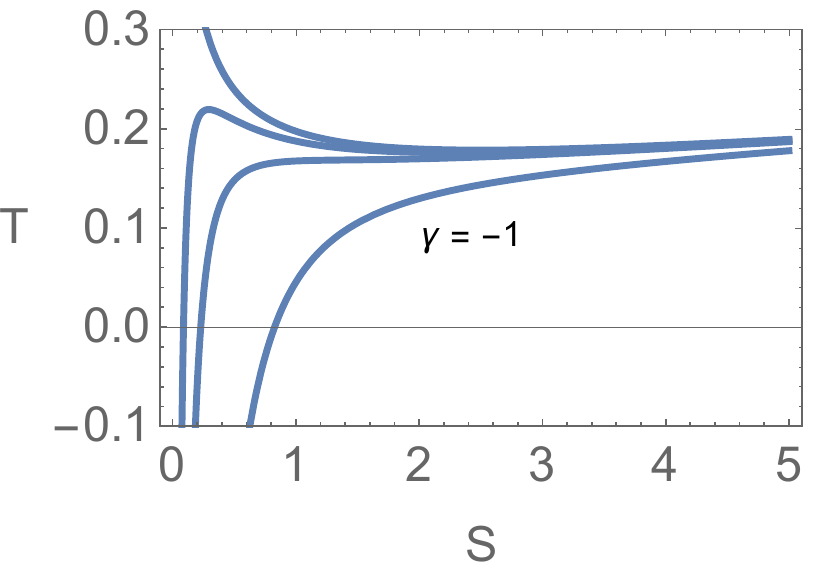}
\caption{The figure shows  $T$ vs $S$ for $ \gamma =-1$. Here  $\lambda = 2.215$, and $ P =  0.05$. The top graph is at $Q=0$ and the rest have $Q = 0.128, 0.222, 0.5$}
\label{temp2}
 \end{center}
 \end{figure}

The temperature of the black hole could  be obtained by using the definition of Hawking temperature: it is based on the surface gravity $\kappa$  at the outer horizon, $r_h$. It is given by,
 \be \label{tempe1}
 T_H =  \frac{ \kappa}{ 2 \pi} = \frac{ 1}{ 4 \pi}  \left| \frac{ df(r)}{ dr} \right|_ { r = r_h} = \frac{1}{ 4 \pi} \left(\frac{ 2 M}{ r_h^2} + \frac{ \gamma Q^2 \lambda}{ r_h^{ \lambda + 1}}   - \frac{ 2 \Lambda r_h}{ 3}\right)
 \ee
 Since  $ f(r_h) = 0$, the mass of the black hole could be written as,
 \be \label{mass}
 M = r_h \left( \frac{1}{2} - \frac{ \gamma Q^2}{ 2 r_h^{( \lambda )} }-  \frac{ r_h^2 \Lambda}{6} \right)
 \ee
 The value of the mass $M$ could  be substituted to  eq$\refb{tempe1}$ and rewrite temperature in terms of $r_h, Q,$ and $\lambda$ as,
 \be \label{temp1}
T = \frac{ 1 }{ 4 \pi} \left( \frac{1}{ r_h}  -  r_h \Lambda   + \frac{ \gamma  ( \lambda -1)Q^2}{ r_h^{ \lambda +1} } \right)
\ee
The conjugate quantity for temperature, the entropy S is given  by the area law, $ S = \pi r_h^2$ for this black hole.
As discussed in the introduction, in the extended phase space, the pressure $P$ is given by,
\be
P = -\frac{ \Lambda}{ 8 \pi}
\ee
The  conjugate quantity for P, the volume $V$,  is given by, $\frac{ 4 \pi r_h^3}{3}$.  The scalar potential which is conjugated to  the scalar charge $Q$ is given by,
\be
\Phi  = - \frac { \gamma Q}{ r_h^{ \lambda -1}}
\ee
One could rewrite the temperature in terms of $S$ and $P$ as follows:
\be
T_H = \frac{1}{ 4 \sqrt{ \pi S}} \left( 1 + 8 P S + \gamma ( \lambda - 1) Q^2 \left(\frac{\pi}{ S}\right)^{\lambda/2} \right)
\ee
The temperature is plotted against the entropy $S$ in Fig.$\refb{temp1}$ and Fig.$\refb{temp2}$. For  $\gamma =1$, the temperature has a minimum: hence  black holes cannot exist below this minimum for a given $\Lambda$ value. For $\gamma =-1$, the temperature could have an inflection point.

\begin{figure} [H]
\begin{center}
\includegraphics{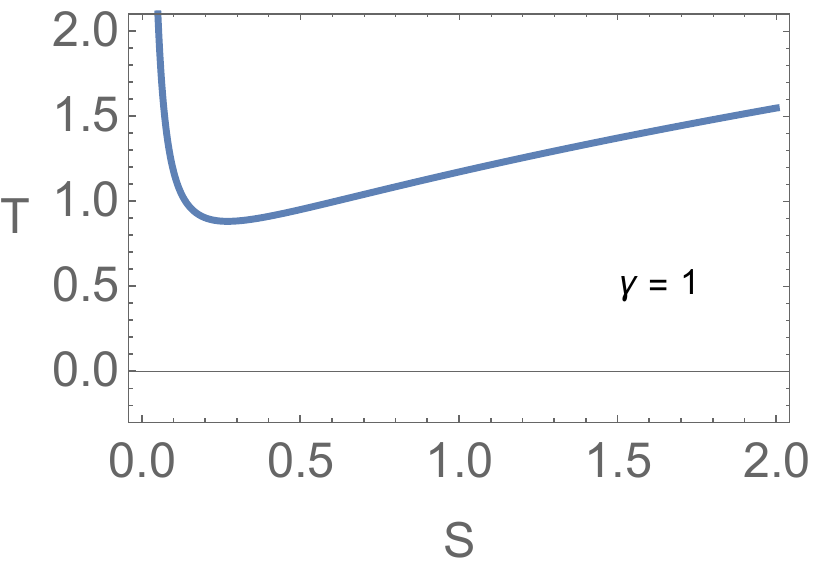}
\caption{The figure shows  $T$ vs $S$ for $ \gamma = 1$. Here  $\lambda = 2.305, P = 0.905$, and $ Q = 0.115$.}
\label{temp1}
 \end{center}
 \end{figure}

\subsection{ First law  in the extended phase space}
 
 In the extended phase space, the mass of the black hole M is not considered as the internal energy as it is usually done. Instead, it is considered as the enthalpy H. Hence, $ M = H = U + P V$.  Hence, the first law of the given black hole is given by,
  \be \label{flaw}
 dM = T dS + \Phi dQ + V dP
 \ee
 It is possible to combine the thermodynamical quantities, $M, P, V, S, T, \Phi$, and $Q$ to obtain the Smarr  formula as,
 \be
 M = 2 T S + \frac{\lambda}{2} \Phi Q - 2 P V
 \ee
which also could  be obtained using the scaling argument presented by Kastor et.al \cite{kastor}. When $\lambda =2$, the Smarr formula simplifies to the one for the Reissner-Nordstrom-AdS black hole obtained by  Kubiznak and Mann in \cite{mann}.


\subsection{ Behavior of pressure}

\begin{figure} [H]
\begin{center}
\includegraphics{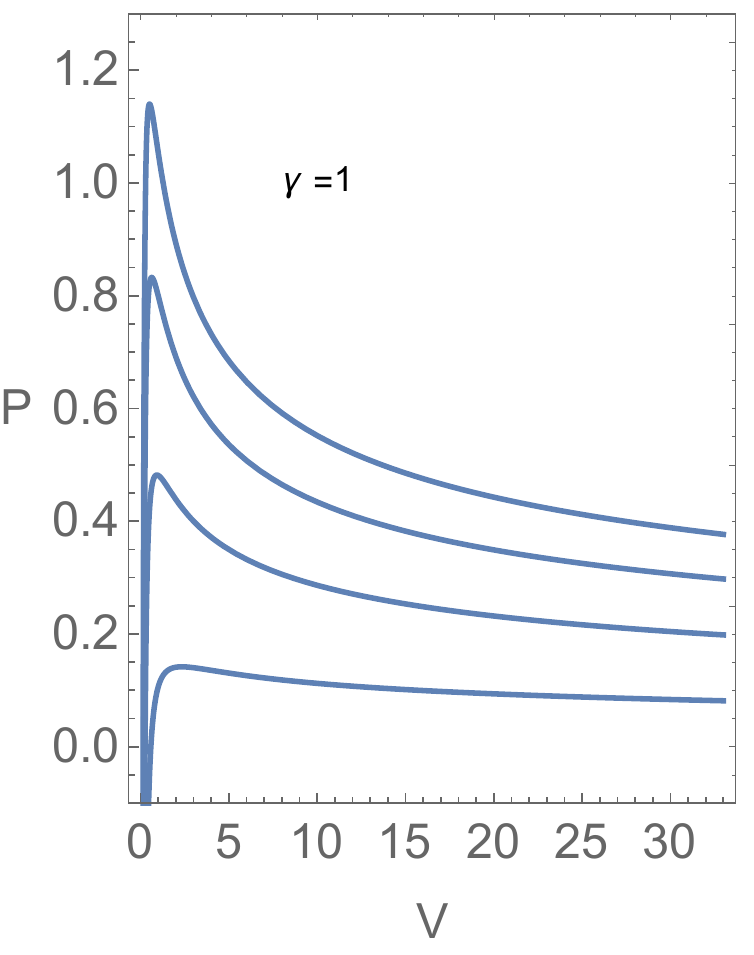}
\caption{The figure shows  $P$ vs $r_h$ for $ \gamma =1$ for varying temperature. Here  $\lambda = 2.865$, and $ Q = 0.334$. For large temperature the peak is higher.}
\label{pvsr2}
 \end{center}
 \end{figure}

\begin{figure} [H]
\begin{center}
\includegraphics{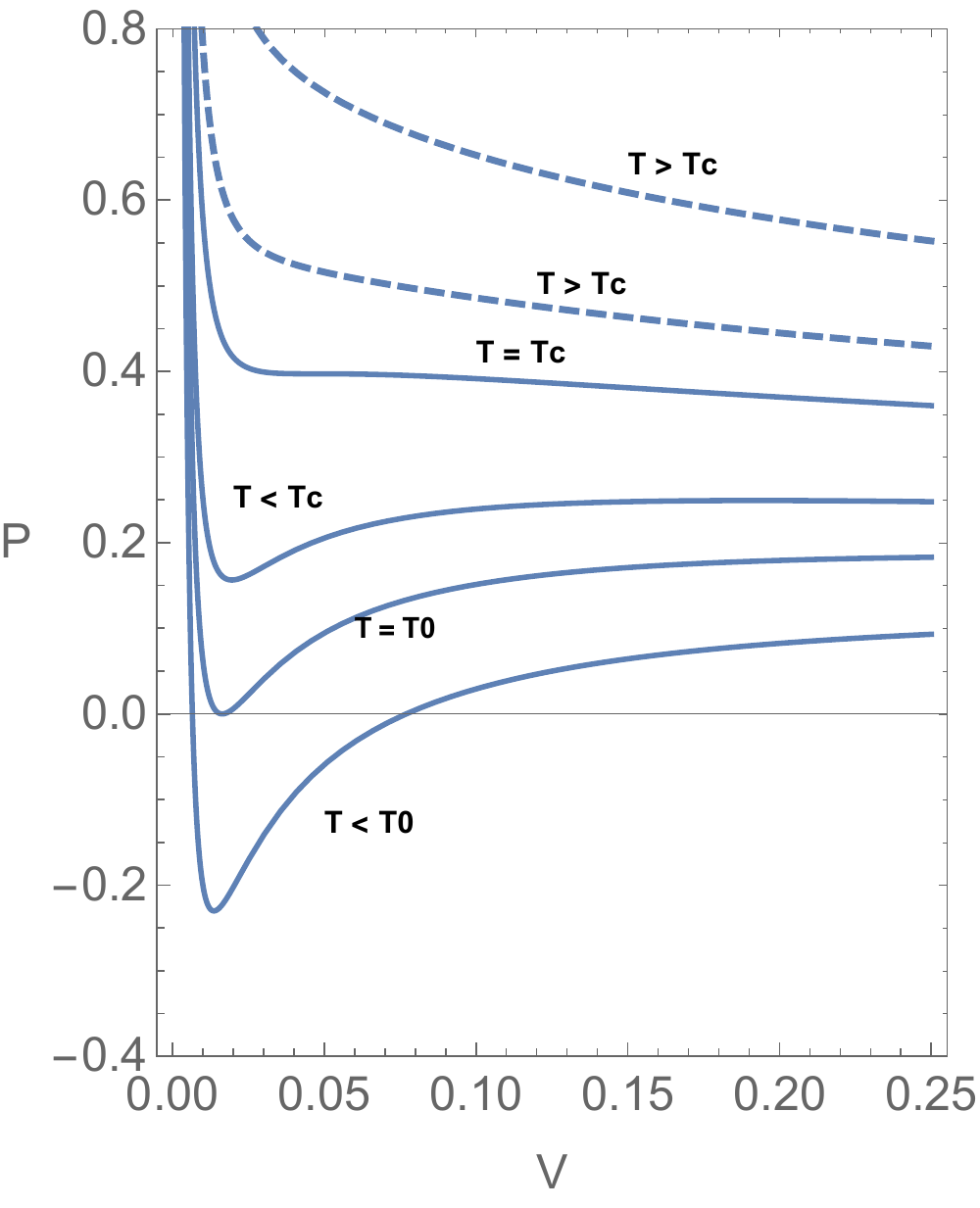}
\caption{The figure shows  $P$ vs $V$ for $ \gamma = -1$ for varying temperature. Here  $\lambda = 1.98$, and $ Q = 0.094$.}
\label{pvsr1}
 \end{center}
 \end{figure}

In eq.$\refb{tempe1}$, one could substitute $ \Lambda = - 8 \pi P$ and rearrange the equation to obtain $P$ as a function of $r_h$ and $T$ as,

\be
P = - \frac{ 1}{ 8 \pi r_h^2}  + \frac{ T}{ 2 r_h} + \frac{ Q^2 \gamma ( 1 - \lambda)}{ 8 \pi r_h^{ 2 + \lambda}}
\ee
Since the black hole radius $r_h$ is given by,
\be
r_h = \left(\frac{ 3 V} {4 \pi}\right)^{1/3}
\ee
one could rewrite $P$ in terms of $V$ as,
\be
P = - \frac{ 1}{ 8 \pi}  \left( \frac{ 4 \pi}{ 3 V} \right) ^ {2/3} + \frac{T}{2} \left( \frac{ 4 \pi}{ 3 V} \right) ^ {1/3} + 
\frac{ Q^2 \gamma ( 1 - \lambda)}{ 8 \pi}  \left( \frac{ 4 \pi}{ 3 V} \right) ^ {(\lambda + 2)/3}
\ee
The pressure is plotted vs V as in Fig.$\refb{pvsr2}$ and Fig.$\refb{pvsr1}$. For $\gamma =1$, there is a maximum pressure given by,
\be \label{pmax}
P_{max} = \frac{ 4 \pi r_h T ( 1 + \lambda) - \lambda}{ 8 \pi r_h^2 ( 2 + \lambda)}
\ee
Since $r_h$ and $V$ are related, $P_{max}$ also could be rewritten in terms of $V_m$ at which the pressure is maximum as,
\be
P_{max} = \frac{ \left( (4 \pi)^{2/3} ( 3 V_m)^{1/3} T ( 1 + \lambda) - \lambda \right)} { 2 ( 2 + \lambda) (4 \pi)^{1/3} ( 3 V_m)^{2/3}}
\ee
Hence when the  $P > P_{max}$ black holes dose not exist. This implies that there is a maximum value of the cosmological constant  that the black holes could exist for a given horizon radius (or volume).

For $\gamma=-1$, the behavior is quite different from what is of $\gamma =1$. There are critical points as demonstrated from Fig$\refb{pvsr1}$.

\subsection{ Phase transitions}

For $\gamma = -1$, there are phase transitions between small and large black holes. These phase transitions are first order and are similar to Van der Waals phase transitions between gas and liquid under constant temperature. A thorough analysis of the phase transitions were discussed in the paper by the current author in \cite{fernando0}.

There is a critical temperature $T_c$ at which the temperature which $P$ vs $V$ curve has an inflection point. At that point,
\be
\frac{ \partial P}{ \partial V } =  \frac{ \partial^2 P}{ \partial V^2 } =0
\ee
The inflection point occur at the volume $V_c$ given by,
\be \label{vc}
V_c = \frac{ 4 \pi r_{hc}^3}{ 3}
\ee
where
\be
r_{hc} =  \frac{1}{2} \left( Q^2 ( \lambda^2 - 1) ( \lambda +2 ) 2 ^{\lambda -1}\right)^{1/\lambda}
\ee
The corresponding $T_c$ and $P_c$ are given by,
\be \label{tc}
T_c = \frac{ \lambda}{ \pi ( \lambda +1)}  \left( Q^2 ( \lambda^2 - 1) ( \lambda +2 ) 2 ^{\lambda -1}\right)^{-1/\lambda}
\ee
\be \label{pc}
P_c = \frac{ \lambda}{ 2 \pi ( \lambda +2)}  \left( Q^2 ( \lambda^2 - 1) ( \lambda +2 ) 2 ^{\lambda -1}\right)^{-2/\lambda}
\ee
 
 
 \subsection{ Specific heat capacities}

There are two different heat capacities for a thermodynamical system: heat capacity at constant pressure, $C_P$, and heat capacity at constant volume, $C_V$. They are given by,
\be
C_P =  T \left.  \frac{\partial S}{ \partial T}\right|_P = \frac{ 2 S \left( 8 P S^{\frac{2 + \lambda}{2}} + S ^{\frac{\lambda}{2}} + \pi^{\frac{\lambda}{2}} Q^2 \gamma ( -1 + \lambda)\right) } { \left( 8 P S^{\frac{2 + \lambda}{2}} - S ^{\frac{\lambda}{2}} - \pi^{\frac{\lambda}{2}} Q^2 \gamma ( -1 + \lambda^2)\right) }
\ee
\be
C_V =  T \left.  \frac{\partial S}{ \partial T}\right|_V = 0
\ee
$C_V$ is zero because the entropy $S$ is proportional to $V$.


\section{ Black hole as a heat engine}

Since the black hole is considered as a thermodynamical system with a $P dV$ term, one could extract mechanical work from the black hole. The given black hole has an equation of state which clearly articulate the relation  between $P$ and $V$. Net work can be extracted from a given cycle in state space. If the net heat input is $Q_H$, net heat out put is $Q_C$ and the net work out put from the system is $W_{net}$, the relation between them are, $Q_H = W_{net} + Q_C$. 

\subsection{Carnot cycle efficiency}

If the given thermodynamical cycle is given by two isothermal and two isentropic paths, then the efficiency of  the heat engine is given by,

\be
\eta_c = \frac{ W_{net}}{ Q_H} = \left( 1 - \frac{ T_C}{T_H} \right)
\ee

\subsection{ Efficiency of a cycle with two isochoric and two isobaric paths}

\begin{figure} [H]
\begin{center}
\includegraphics{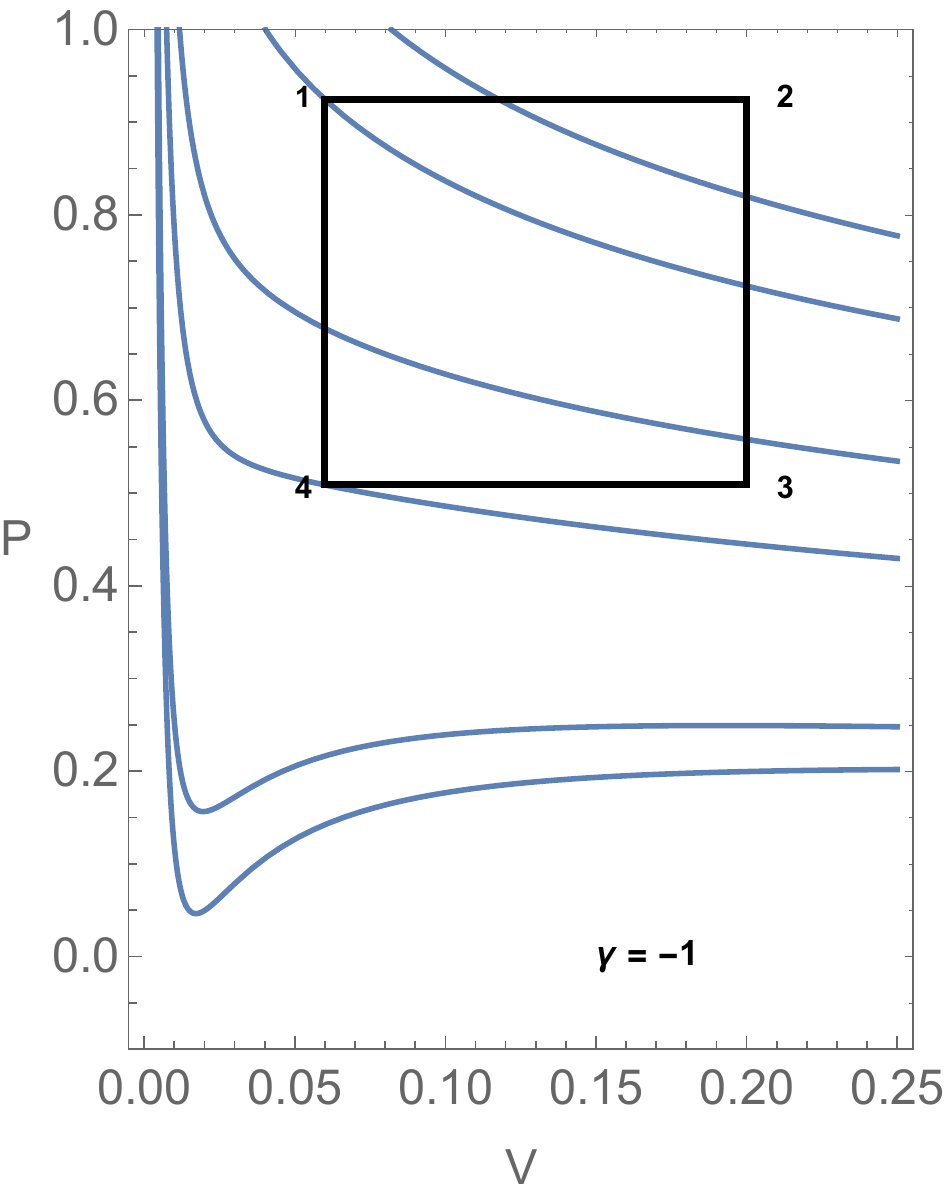}
\caption{The figure shows  $P$ vs $V$ for $ \gamma = -1$ for varying temperature. Here  $\lambda = 1.98$, and $ Q = 0.094$. The maximum pressure, $P_1 = P_2 = 0.925$ and minimum pressure, $P_3 = P_4 = 0.503$.}
\label{pvcycle}
 \end{center}
 \end{figure}

In this paper, we focus on a thermodynamical path  consisting of two constant pressure and two constant volume paths as shown in Fig.$\refb{pvcycle}$. The heat supplied along the isobaric path $1 \ra 2$ which is the higher temperature path is given by,
\be \label{heat1}
Q_{1 \ra 2} = Q_H = \int_1^2 d Q = \int_1^2 T dS
\ee
Since $S = \pi r_h^2$,  eq.$\refb{heat1}$ can be rewritten as,
\be \label{integral}
Q_H = \int_1^2 T_H d ( \pi r_h^2) = 2 \pi \int_{r_1}^{r_2} T_H(r_h) r_h dr_h
\ee
Since the pressure $P$  is constant along the path $1 \ra 2$, and $Q, \lambda$ are also kept constant, $T_H$ becomes a function of $r_h$ along $1 \ra 2$. Hence the integral in eq.$\refb{integral}$ could be computed to be,
\be
Q_H =   \left( \frac{r_h}{2} - \frac{ \gamma Q^2}{ 2 r_h^{( \lambda -1 )} }-  \frac{ 4 \pi r_h^3  P}{3} \right)^{r_2}_{r_1} = M(r_2) - M(r_1)
\ee
Similarly the net heat out put from the system could be calculated along the path $4 \ra 3$ as,
\be
Q_{ 3 \ra 4} = Q_{C} = M(r_3) - M(r_4)
\ee
Therefore, the efficiency of the the heat engine for this particular cycle is,
\be \label{eta}
\eta = \frac{W_{net}}{ Q_H} = 1 - \frac{ (M(r_3) - M(r_4) )}{ (M(r_2) - M(r_1) )}
\ee
The above formula was obtained by Johnson \cite{johnson2} by using enthalpy and the first law of thermodynamics.

It is also noted that since volume is a function of entropy in this black hole, isochoric path also represents adiabatic path.

\section{Efficiency of the heat engine  for varying parameters of the black hole}

The goal of this  section is to compute the efficiency of the black hole heat engine and compare to the Carnot efficiency $\eta_C$.  We will restrict the black holes with $\gamma = -1$ for the rest  of the paper. There are two parts to this computation: first we will change $Q$ for three values and perform the calculations, and second we will change $\lambda$ and perform the computations. In both cases,  $V$ (or $r_h$) and the $P$  for the coordinates 1 and 4 are kept constant. Then $V$ (or $r_h$) for coordinates  2 and 3 are varied to observe how the efficiency varies. 

\begin{figure} [H]
\begin{center}
\includegraphics{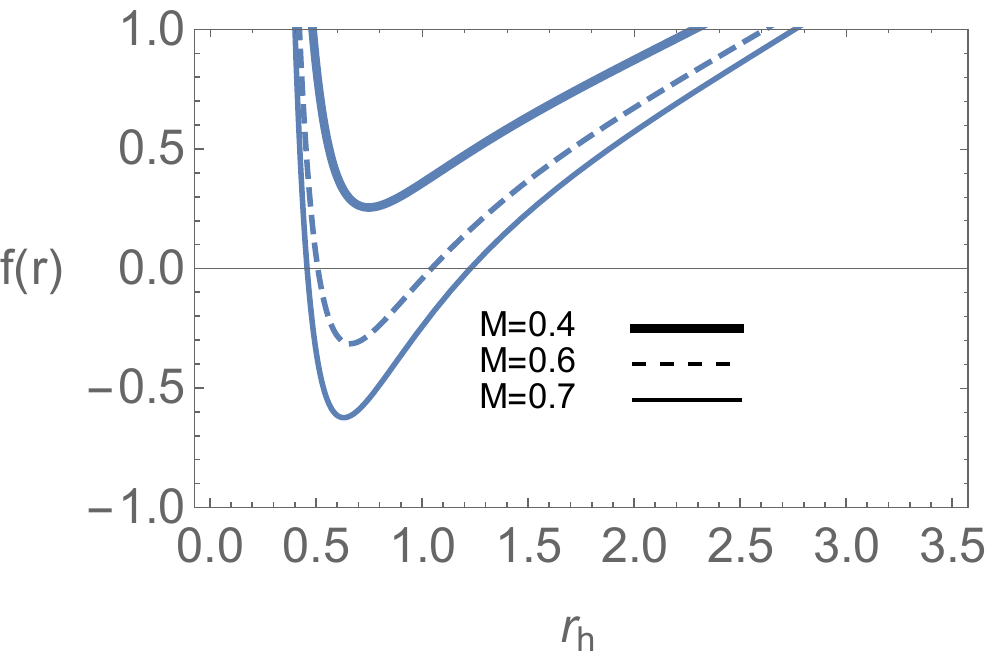}
\caption{The figure shows  $f(r)$ vs $r$ for $ \gamma = -1$ for varying mass. Here  $\lambda = 4, \Lambda = -0.2$, and $ Q = 0.3$. }\label{frvsmass}
 \end{center}
 \end{figure}

Before we proceed,  some cautionary remarks are in order:  for a given mass, the horizon radius $r_h$ is possible only if the parameters of the black hole are chosen appropriately. To clarify this further we have shown an example in the Fig.$\refb{frvsmass}$ of the function $f(r)$ whose roots lead to the horizon radius $r_h$. In this example, when the mass is low, there are no horizons. Only for a mass greater than a critical value that the horizon radius exists.

\begin{figure} [H]
\begin{center}
\includegraphics{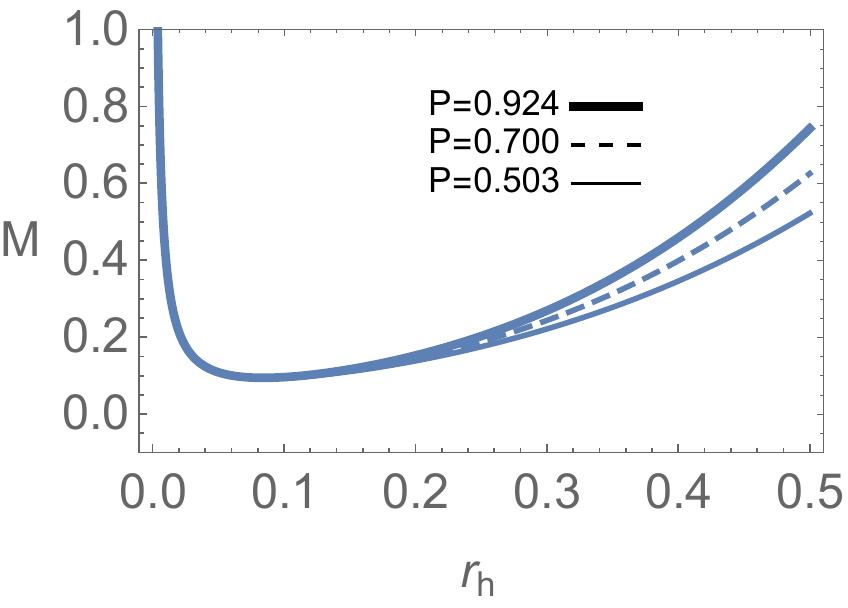}
\caption{The figure shows  $M$ vs $r_h$ for $ \gamma = -1$ for varying pressure. Here  $\lambda = 1.98$, and $ Q = 0.094$. The values of the pressures are chosen so that they are between the maximum and the minimum  pressures of the cycle, $P_1 = P_2 = 0.925$ and $P_3 = P_4 = 0.503$ respectively.}
\label{mvsr}
 \end{center}
 \end{figure}

On the other hand, we also would like to make sure that for all the points on the chosen thermodynamical cycle that there is a real mass value. In Fig.$\refb{mvsr}$,  $M$ is plotted vs $r_h$ for various values of $P$. The values of the pressures are chosen so that they are between the maximum and the minimum pressure of the cycle considered. It is clear that the mass $M$ is positive for all values of $P$ chosen. However, it is interesting to note that the mass has a minimum value for a given $P$.  We did a spot check to make sure that for the range of $r_2$ chosen, that $M(r_2), M(r_4) > 0$.

\subsection{Efficiency for varying scalar charge $Q$}

\begin{figure} [H]
\begin{center}
\includegraphics{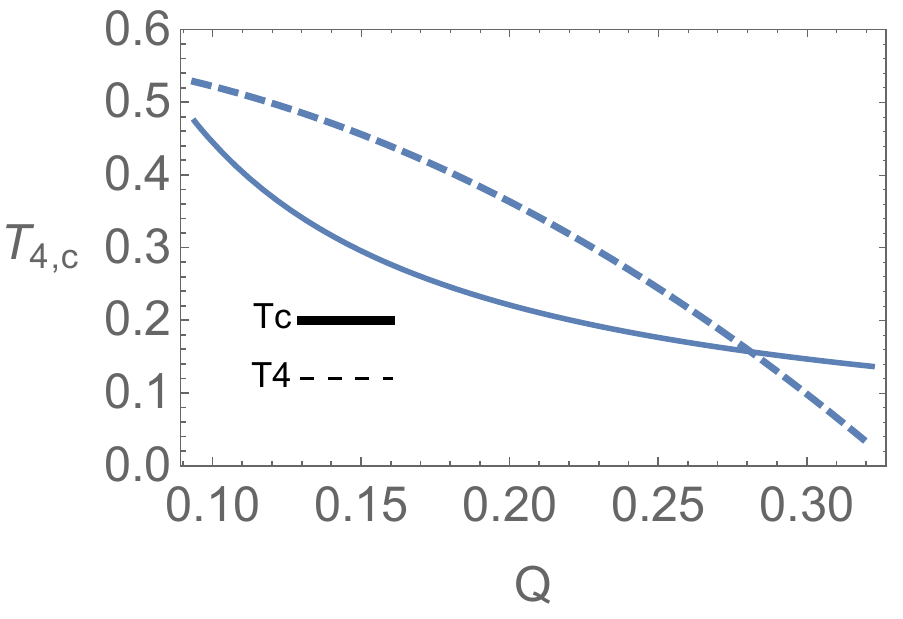}
\caption{The figure shows  $T_c, T_4$ vs $Q$ for $ \gamma = -1$. Here  $\lambda = 1.98$, and $P_4 = 0.509$. }
\label{tct4}
\end{center}
\end{figure}

In this section, the scalar charge $Q$ will be varied for three values. Here, $P_1, P_4, r_4= r_1$ will be kept constant. In the thermodynamical cycle in Fig.$\refb{pvcycle}$, the smallest temperature is at $T_4$. The cycle is chosen such that it does not include the  region where the  phase transition occur. Due to this reason, the critical temperature $T_c$ needs to be smaller than $T_4$ for all chosen $Q$ values. $T_c$ is computed by eq.$\refb{tc}$. $T_4$ is computed as,
\be \label{temp4}
T_4 = \frac{ 1 }{ 4 \pi} \left( \frac{1}{ r_4}  + 8 \pi P_4 r_4   + \frac{ \gamma  ( \lambda -1)Q^2}{ r_4^{ \lambda +1} } \right)
\ee
$T_c$ and $T_4$ are plotted together in the same graph in Fig.$\refb{tct4}$ for varying $Q$. For large $Q$, $T_c > T_4$. Hence one need to choose the correct range of $Q$ values to make sure that $T_c <  T_4$.

\begin{figure} [H]
\begin{center}
\includegraphics{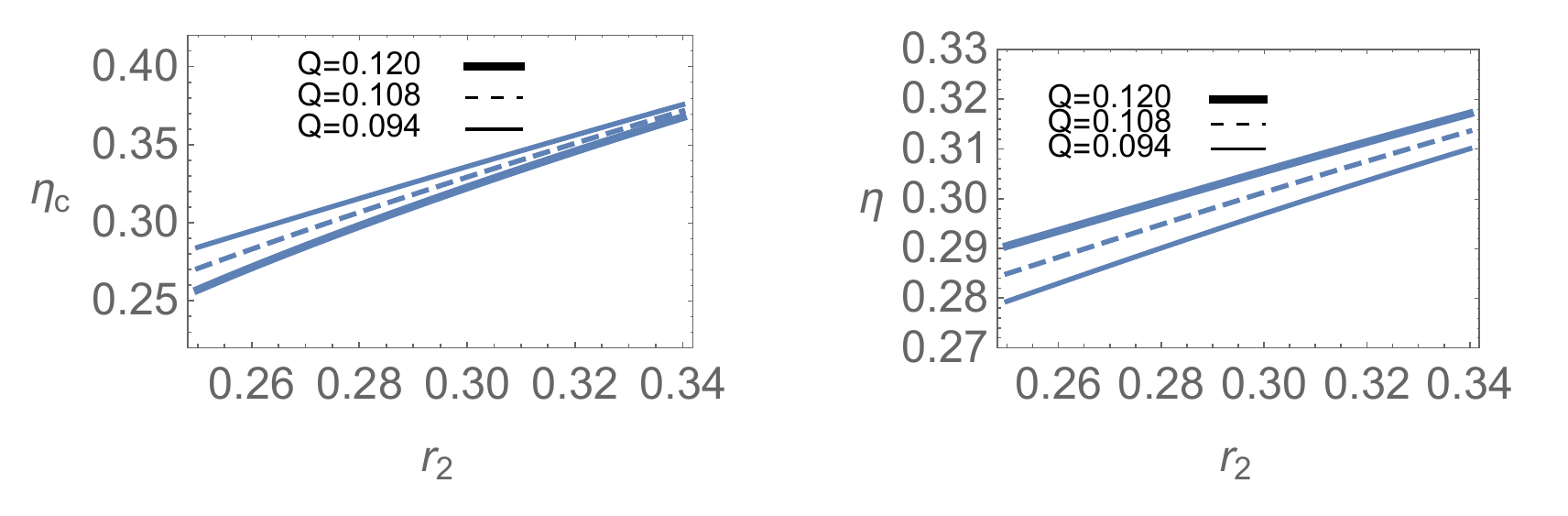}
\caption{The figure shows  $\eta, \eta_c$ vs $r_2$ for $ \gamma = -1$ for varying scalar charge $Q$. Here  $\lambda = 1.98$}
\label{etavsq}
\end{center}
\end{figure}

When $Q$ is fixed, $M(r_3)$ and $M(r_2)$ will depend only on $r_3 = r_2, P_1$ and $P_4$. Now, $\eta, \eta_C$ are calculated using eq.$\refb{eta}$ and plotted for 3 different values of $Q$ in Fig.$\refb{etavsq}$.

\begin{figure} [H]
\begin{center}
\includegraphics{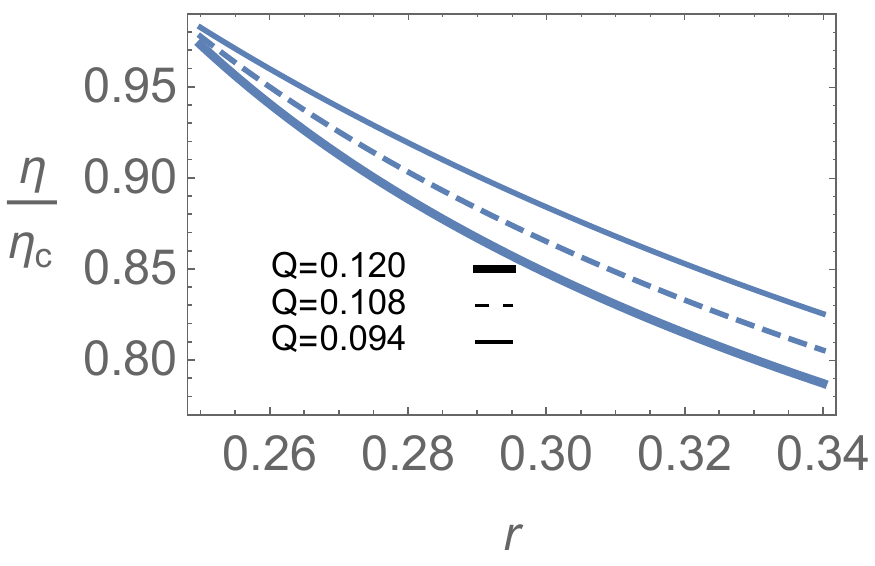}
\caption{The figure shows  $\frac{\eta}{\eta_c}$ vs $r_2$ for $ \gamma = -1$ for varying scalar charge $Q$. Here  $\lambda = 1.98$}
\label{ratioq}
\end{center}
\end{figure}

\subsection{Efficiency for varying  $\lambda$}

\begin{figure} [H]
\begin{center}
\includegraphics{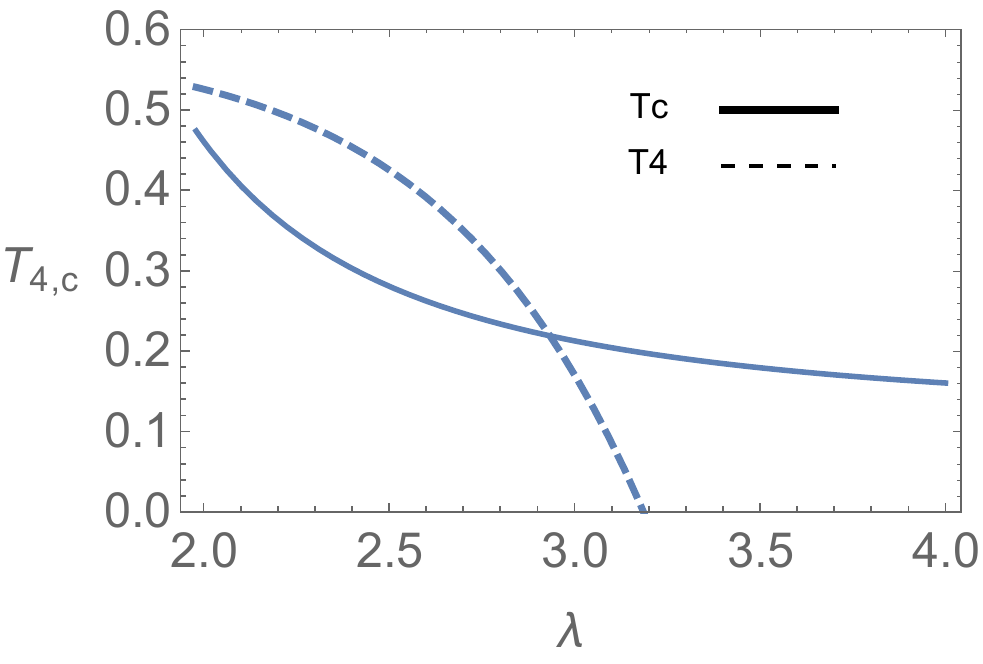}
\caption{The figure shows  $T_c, T_4$ vs $\lambda$ for $ \gamma = -1$. Here  $Q = 0.094$, and $P_4 = 0.509$. }
\label{tct4lambda}
\end{center}
\end{figure}

In this section, the value of $\lambda$ is varied for three values. Once again, $P_1, P_4, r_4= r_1$ will be kept constant. It is also important to make sure that $T_c < T_4$ for chosen values of $\lambda$. In Fig.$\refb{tct4lambda}$, $T_c, T_4$ are plotted against $\lambda$. It is clear that there is a value $\lambda$ that $T_c > T_4$. Hence values of $\lambda$ are chosen such that $T_c < T_4$.

\begin{figure} [H]
\begin{center}
\includegraphics{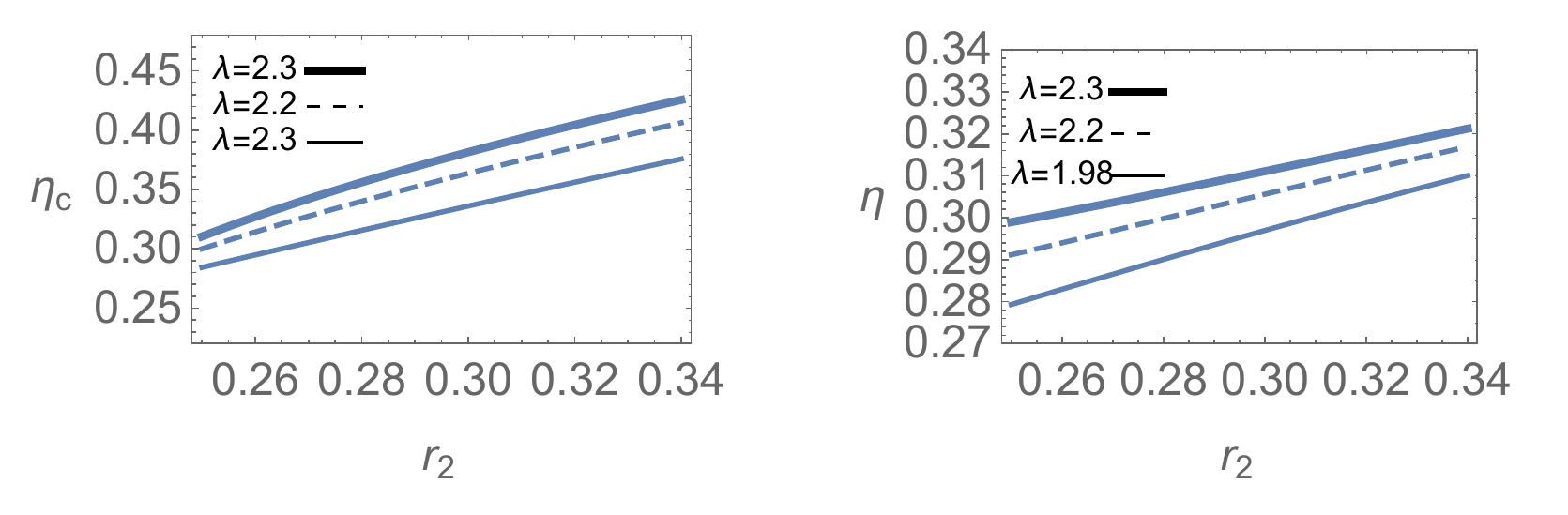}
\caption{The figure shows  $\eta, \eta_c$ vs $r_2$ for $ \gamma = -1$ for varying $\lambda$. Here  $Q = 0.094$}
\label{etalambda}
\end{center}
\end{figure}

\begin{figure} [H]
\begin{center}
\includegraphics{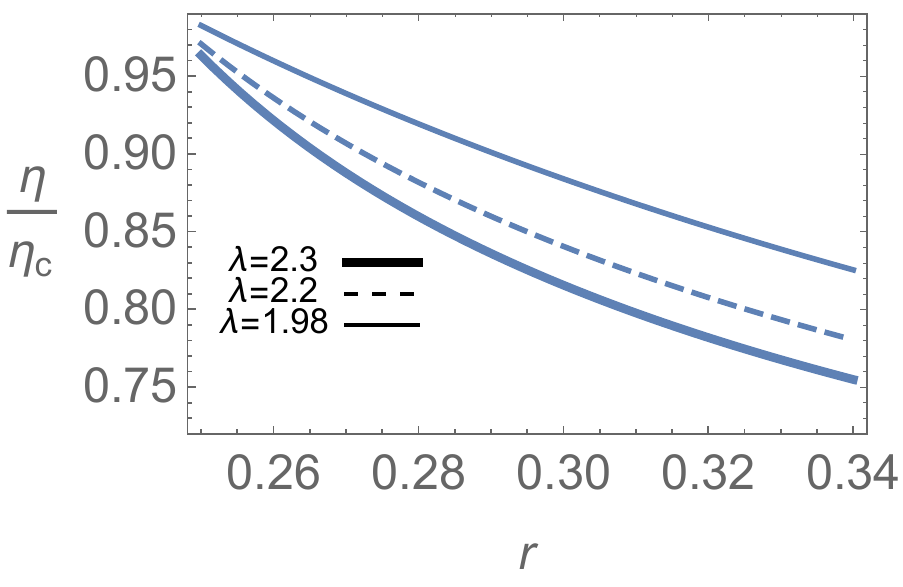}
\caption{The figure shows  $\frac{\eta}{\eta_c}$ vs $r_2$ for $ \gamma = -1$ for varying $\lambda$. Here  $Q = 0.094$}
\label{ratiolambda}
\end{center}
\end{figure}

\section{ Conclusion}

In this paper, we have studied black holes in massive gravity with a negative cosmological constant. in the extended phase space. Here the black hole has a pressure given by,  $P = -\frac{\Lambda}{ 8 \pi}$. There are two  types of black holes for the values of $\gamma$ in the theory. 
Thermodynamical behavior differ significantly for $\gamma =1$ and $\gamma =-1$. For $\gamma =1$, the  black hole behaves similar to the Schwarzschild-anti-de Sitter black hole: the pressure has a maximum and the temperature has a minimum.  For $\gamma =-1$, the black hole exhibits phase transitions for certain range of temperatures: for higher temperatures, the black holes behave like an ideal gas. Phase transitions are between large black holes and small black holes. A detailed analysis of the black holes in this context is published by the current author in \cite{fernando0}.

The main goal of this paper  is to study the massive gravity black hole as a heat engine. The thermodynamical cycle for the heat engine considered here is a rectangle in $P, V$ space with two isobaric  and two isochoric processes. The efficiency of the heat engine taking black hole as the working substance is computed for the rectangle cycle as well as for the Carnot cycle by varying $Q, \lambda, r_2$. When $Q$ is increased, the efficiency for the rectangle cycle increases, but, the Carnot efficiency decreases. The ratio $\frac{\eta}{\eta_c}$ decreases when $Q$ is increased. Hence to achieve better efficiency, a higher $Q$ is appropriate. Larger the volume $V_2$ (or $r_2$), higher the $\eta$ of the rectangle cycle. When $Q =0$, we get the Schwarzschild AdS black hole. Hence from the graphs it can be concluded that the Schwarzschild AdS black hole has a smaller efficiency for the rectangle cycle compared with the massive gravity black hole.

When $\lambda$ is increased, the efficiency of the rectangle cycle as the Carnot cycle increases. Both efficiencies increase with  volume $V_2$ (or $r_2$). The ratio $\eta/\eta_c$ decreases with $\lambda$.

The Reissner-Nordstrom AdS (RNAdS) black hole has the metric with 
\newline
$f(r) = 1 - \frac{2 M}{r} + \frac{Q_e^2}{r^2} - \Lambda r^2$. When $Q_e = Q$ and $\lambda =2$, the massive gravity black hole and the RNAdS black hole are the same. Hence one can conclude that for $\lambda >2$ with  $ Q_e = Q$, the massive gravity black hole will have higher efficiency. When $ 1 \leq \lambda < 2 $, massive gravity black hole  will have lower efficiency.

\vspace{0.5 cm}



\end{document}